\documentclass[twocolumn,showpacs,preprintnumbers,amsmath,amssymb]{revtex4}

\usepackage{graphicx}
\usepackage{dcolumn}
\usepackage{bm}

\begin{document}

\preprint{APS/123-QED}

\title{The Effect of the Pauli Exclusion Principle in the Many-Electron Wigner Function}%

\author{Emiliano Cancellieri, Paolo Bordone}
\author{Carlo Jacoboni}%
\affiliation{Dipartimento di Fisica Universit\`a di Modena e Reggio Emilia and
CNR-INFM S3 National Research Center, via Campi 213/a, 41100 Modena (Italy).}

\date{\today}

 \begin{abstract}
 An analysis of the Wigner function for identical particles is presented. Four situations have been
 considered. i) A scattering process between two indistinguishable electrons described by a minimum
 uncertainty wave packets showing the exchange and correlation hole in Wigner phase space. ii) An
 equilibrium ensemble of N electrons in a one-dimensional box and in a one-dimensional harmonic
 potential showing that the reduced single particle Wigner function as a function of the energy
 defined in the Wigner phase-space tends to a Fermi distribution. iii) The reduced one-particle
 transport-equation for the Wigner function in the case of interacting electrons showing the need
 for the two-particle reduced Wigner function within the BBGKY hierarchy scheme. iv) The
 electron-phonon interaction in the two-particle case showing co-participation of two electrons in
 the interaction with the phonon bath.
 \end{abstract}

\pacs{05.30.Fk; 63.20.-e; 72.10.-d}
\maketitle

 \section{Introduction}

Highly sophisticated technologies produce physical systems, and in particular semiconductor
devices, of very small dimensions, comparable with electron wavelength or with electron coherence
lengths. Under such conditions, semi-classical dynamics is not justified in principle and
interference effects due to the linear superpositions of quantum states have to be considered.
Among the possible different approaches, the Wigner-function (WF) has proved to be very useful
for studying quantum electron transport \cite{Wigner,Frensley,Nedjalkov,Jacoboni1}, owing to
its strong analogy with the semiclassical picture, since it explicitly refers to variables
defined in an $({\bf r},{\bf p})$ Wigner phase space, together with a rigorous description of
electron dynamics in quantum terms.

In this work we present an analysis of the WF for identical particles. Even thought the WF was
defined from its very beginning for the study of many-particle physics, in electron transport
theory it has been used mainly in its one-particle version. The importance of the many-body problem
derives from the fact that any real physical system one can think of is composed of a set of
interacting bodies. Moreover, since we are dealing with quantum mechanical systems the symmetry
properties that describe the behavior of identical particles play an essential role. The present
paper will be focused mainly on the last subject.

In particular, four situations will be analyzed: i) A scattering process between two
indistinguishable electrons described by minimum uncertainty wave packets, showing the exchange
and correlation hole in Wigner phase space. ii) An equilibrium ensemble of N electrons in a
box and in a harmonic potential, showing that the sum of the values of the WF that correspond
to points in the Wigner phase-space with energy in a given interval, tends to a
Fermi distribution. iii) The transport equation for interacting electrons, showing the BBGKY
hierarchy when the integral, over the degrees of freedom of all the particles but one, are
performed \cite{Carruters,Rossi}. iv) The electron-phonon interaction in the case of two
particles, where new Keldysh diagrams \cite{Keldysh} appear with respect to the one-electron
case \cite{Pascoli}.

\section{Wigner Function for Many Identical Particles}

The WF was introduced by Wigner in 1932 to study quantum corrections to classical statistical
mechanics \cite{Wigner,Uhlenbeck,Kirkwood,Harper}. Thus, even though it is now used mainly in
single particle problems, from the very beginning this function was defined for N particles as:

\begin{eqnarray}
\label{defWF}
f_{W}({\bf r}_1,{\bf p}_1,...,{\bf r}_N,{\bf p}_N,t)
& = &
\int\,d{\bf s}_1...d{\bf s}_Ne^{-\frac{i}{\hbar}\sum{\bf s}_i{\bf p}_i}
\nonumber\\
&&
\hspace{-0.35cm}
\times
\psi\left({\bf r}_1+\frac{{\bf s}_1}{2},...,{\bf r}_N+\frac{{\bf s}_N}{2},t\right)
\nonumber\\
&&
\hspace{-0.35cm}
\times
\psi^{\star}
\hspace{-0.08cm}
\left({\bf r}_1-\frac{{\bf s}_1}{2},...,{\bf r}_N-\frac{{\bf s}_N}{2},t\right)
\hspace{-0.08cm}.
\nonumber\\
\end{eqnarray}

In the case of identical particles, the wave function describing the many-body system satisfy
well known symmetry relations. When the position coordinates of two particles are interchanged
the wave function remains unaffected (bosons) or changes sign (fermions).  Since the WF is bilinear
in the wave function it remains the same if the positions and, accordingly, the Wigner momenta of two
particles are exchanged.

This symmetry property of the WF allows the definition of a reduced $M$-particle WF in a system
of $N$ particles as \cite{Hillery,Imre}:

\begin{widetext}
\begin{equation}
\label{reduced}
f^{(N)}_{W}({\bf r}_1,{\bf p}_1,...,{\bf r}_M,{\bf p}_M,t)
=
\frac{N!}{(N-M)!h^{3(N-M)}}
\int\,d{\bf r}_{M+1}\,d{\bf p}_{M+1}...d{\bf r}_N\,d{\bf p}_N
f_{W}({\bf r}_1,{\bf p}_1,...,{\bf r}_N,{\bf p}_N,t),
\end{equation}
\end{widetext}

\noindent
where the superscript $(N)$ indicates that the reduced $M$-particle WF is defined in a system
with $N$ particles. Note that in the case where $M=1$ the above equation becomes:

\begin{eqnarray}
\label{reducedtoone}
f^{(N)}_{W}({\bf r}_1,{\bf p}_1,t)
& = &
\frac{N}{h^{3(N-1)}}\int\,d{\bf r}_{2}\,d{\bf p}_{2}...d{\bf r}_N\,d{\bf p}_N
\nonumber
\\
&&
\times
f_{W}({\bf r}_1,{\bf p}_1,...,{\bf r}_N,{\bf p}_N,t).
\nonumber
\\
\end{eqnarray}

\noindent
The factorials appearing in front of the integral in equation (\ref{reduced}) simplify to $N$ in
equation (\ref{reducedtoone}) since this is the number of equivalent ways one can reduce the
$N$-particle WF when the particles themself are supposed to be identical.

\subsection{The WF for Many Single-Particle Wave Functions}

We consider the case of $N$ particles in the system and we define the WF with a wave function
that is a symmetric or anti-symmetric linear combination of single-particle wave functions
$\psi_i({\bf r})$, $(i=1,...,N)$ as

\begin{eqnarray}
\psi({\bf r}_1,...,{\bf r}_N)
& = &
\psi_1({\bf r}_1)\psi_2({\bf r}_2)\psi_3({\bf r}_3)\,...\,\psi_N({\bf r}_N)
\nonumber
\\
&&
\pm
\psi_1({\bf r}_2)\psi_2({\bf r}_1)\psi_3({\bf r}_3)\,...\,\psi_N({\bf r}_N)
\nonumber
\\
&&
+\psi_1({\bf r}_2)\psi_2({\bf r}_3)\psi_3({\bf r}_1)\,...\,\psi_N({\bf r}_N)
\nonumber
\\
&&
\pm
\psi_1({\bf r}_3)\psi_2({\bf r}_2)\psi_3({\bf r}_1)\,...\,\psi_N({\bf r}_N)
\nonumber
\\
&&
+\,...
\end{eqnarray}

\noindent
where the upper sign is for bosons and the lower for fermions. In the WF expression it is possible
to identify two different types of terms. The first one is characterized by the product of single-particle
WFs. In each of these contributions, from the different wave functions, $N$ WF are obtained that
are evaluated in a particular permutation of the variable indices as, for example: 
$f_{W_1}({\bf r}_4,{\bf p}_4)f_{W_2}({\bf r}_1,{\bf p}_1)f_{W_3}({\bf r}_2,{\bf p}_2)...f_{W_N}({\bf r}_{N-5},{\bf p}_{N-5})$.

The second type of contributions accounts for the exchange effects and vanishes when the wave
functions $\psi_n({\bf r})$ do not overlap. These terms are constituted by integrals of the product
of $N$ factors $\psi_n\,\psi^{\star}_n$, one for each of the $N$ wavefunctions $\psi_n$. In these
terms at list two products $\psi_n({\bf r}_i+{\bf s}_i/2)\psi^{\star}_n({\bf r}_j-{\bf s}_j/2)$
are evaluated with $i\ne j$. It is the presence of such factors that makes impossible to obtain the
many-particle WF in terms of single-particle WFs. The number of factors $\psi_n\,\psi^{\star}_n$,
where $\psi_n$ and $\psi^{\star}_n$ correspond to different particles, appearing in a given integral
can range from $2$ to $N$. As an example the WF in the case of $N=2$ reads:

\begin{widetext}
\begin{eqnarray}
\label{overlapping}
f_W({\bf r}_1,{\bf p}_1,{\bf r}_2,{\bf p}_2,)
& = &
f_{W1}({\bf r}_1,{\bf p}_1)f_{W2}({\bf r}_2,{\bf p}_2)+
f_{W1}({\bf r}_2,{\bf p}_2)f_{W2}({\bf r}_1,{\bf p}_1)
\nonumber\\
&&
\pm\frac{1}{\hbar^6}\int\,d{\bf s}_1\,d{\bf s}_2
e^{-\frac{i}{\hbar}({\bf s}_1{\bf p}_1+{\bf s}_2{\bf p}_2)}
\nonumber\\
&&
\times
\left[
\psi_1\left({\bf r}_1+\frac{{\bf s}_1}{2}\right)
\psi^{\star}_1\left({\bf r}_2-\frac{{\bf s}_2}{2}\right)
\psi_2\left({\bf r}_2+\frac{{\bf s}_2}{2}\right)
\psi^{\star}_2\left({\bf r}_1-\frac{{\bf s}_1}{2}\right)
\right.
\nonumber\\
&&
+
\left.
\psi_1\left({\bf r}_2+\frac{{\bf s}_2}{2}\right)
\psi^{\star}_1\left({\bf r}_1-\frac{{\bf s}_1}{2}\right)
\psi_2\left({\bf r}_1+\frac{{\bf s}_1}{2}\right)
\psi^{\star}_2\left({\bf r}_2-\frac{{\bf s}_2}{2}\right)
\right],
\nonumber\\
\end{eqnarray}
\end{widetext}

\noindent
here $4$ terms appear, $2$ for each kind of contribution. The two-particle system is treated in
details in \cite{iwce10}.

\subsection{Example of Two Colliding Electrons}

\begin{figure}
\includegraphics[width=7.5cm]{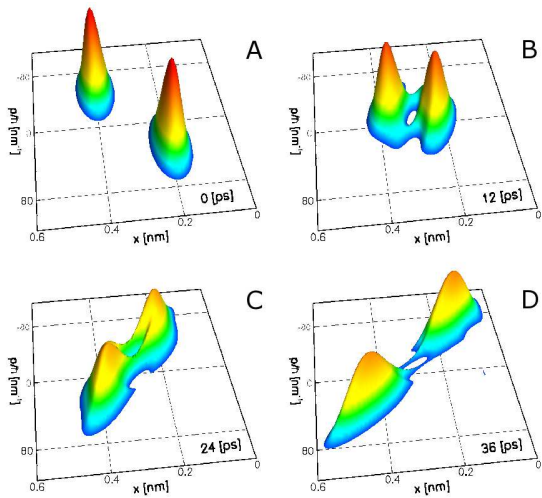}
\caption{\label{fourWF} One dimensional reduced single-particle WF of two interacting electrons
at different times. The figure clearly shows the exchange hole due to the Pauli exclusion principle.}
\end{figure}

A one-dimensional situation where two fermions collide with each other has been simulated.
The Schr\"odinger equation was solved with initial conditions given by two minimum-uncertainty
wave packets interacting through the Coulomb potential and the WF was evaluated at different time
steps (see Fig.~\ref{fourWF}).

In this figure we plot the one-particle reduced WF of the system for the case of two Gaussian
wave packets with opposite central wave vectors. Since we are dealing with a one dimensional
system, the two particles are expected to decelerate, scatter, and then move away from
each other. At $t=0$ we suppose the two particles to be described by an antisymmetric wave
function. In Fig. $1.B$ the system is shown $2$ ps after the Coulomb interaction is
switched on. At the beginning of the scattering process the exchange hole due to the Pauli's
exclusion principle appears. In part $1.C$ the two particles are shown when their mutual
distance has reached the minimum value. In this case the exchange hole is maximally evident.
When the two particles are moving far enough from each other the exchange hole tends to disappear
(part $1.D$).

\section{Equilibrium WF for non Interacting Particles in Confined Potentials}

In this section a system of $N$ fermions in a confined potential has been studied. In the
non-interacting case the one-particle reduced WF is studied at thermal equilibrium at a temperature
of $T=2$ K. In order to simplfy the mathematical treatment we shall introduce the second quantization
notation. The $N$-particles wave function can thus be written as:

\begin{equation}
\psi({\bf r}_1,...,{\bf r}_N)=\langle{\bf r}_1,...,{\bf r}_N|\psi\rangle
=\langle 0 |\hat{\Psi}({\bf r}_1)...\hat{\Psi}({\bf r}_N)|\psi\rangle,
\end{equation}

\noindent
and the WF as:

\begin{widetext}
\begin{eqnarray}
\label{defsecond}
f^{(N)}_W({\bf r}_1,{\bf p}_1)
& = &
\frac{N}{h^{3(N-1)}}\int\,d{\bf r}_{2}\,d{\bf p}_{2}...d{\bf r}_N\,d{\bf p}_N
\int\,d{\bf s}_{1}...d{\bf s}_N
e^{-\frac{i}{\hbar}\sum_{j=1}^N{\bf p}_j{\bf s}_j}\times
\nonumber
\\
&&
\Big\langle 0\Big|\hat{\Psi}\left({\bf r}_1+\frac{{\bf s}_1}{2}\right)...
\hat{\Psi}\left({\bf r}_N+\frac{{\bf s}_N}{2}\right)
{\overline{\Big|\psi\Big\rangle\Big\langle\psi\Big|}}
\hat{\Psi}^{\dagger}\left({\bf r}_N-\frac{{\bf s}_N}{2}\right)...
\hat{\Psi}^{\dagger}\left({\bf r}_1-\frac{{\bf s}_1}{2}\right)\Big|0\Big\rangle,
\nonumber
\\
\end{eqnarray}
\end{widetext}

\noindent
where, here and in the following, $\hat{\Psi}$ and $\hat{\Psi}^{\dagger}$ are the creation and
annihilation field operators. Since we are interested in the thermal equilibrium distribution
of a fixed number of particles, the density matrix in the above equation is:

\begin{equation}
\overline{|\psi\rangle\langle\psi|}=\hat{\rho}=\frac{1}{\mathcal{Z}}e^{-\frac{\hat{H}}{k_BT}},
\end{equation}

\noindent
where $\mathcal{Z}$ is the partition function, $\hat{H}$ the Hamiltonian, $k_B$ the Boltzman
constant, and $T$ the temperature of the system. Since the particles in the system are supposed
to be non interacting, and the system is supposed to be confined, the Hamiltonian in its second
quantization form can easely be written in terms of the particles creation $\hat{c}^{\dagger}_n$ and
annihilation operators $\hat{c}_n$ as:

\begin{equation}
\hat{H}=\sum_{n=1}^{\infty}\epsilon_n\hat{c}^{\dagger}_n\hat{c}_n,
\end{equation}

\noindent
where $\epsilon$ is the energy of the $n$-th discrete level. Writing the field operators in terms of
the creation and annihilation operators, the mean value element appearing in equation (\ref{defsecond})
becames:

\begin{widetext}
\begin{eqnarray}
\label{matrixelemnet}
\langle 0|...|0\rangle
& = &
\sum_{n'_1}...\sum_{n'_N}\sum_{n''_1}...\sum_{n''_N}
\left\langle 0\left|\hat{c}_{n'_1}...\hat{c}_{n'_N}
e^{-\frac{1}{k_BT}\sum_n\epsilon_n\hat{c}^{\dagger}_n\hat{c}_n}
\hat{c}^{\dagger}_{n''_N}...\hat{c}^{\dagger}_{n''_1}\right|0\right\rangle\times
\nonumber
\\
&&
\psi_{n'_1}\left({\bf r}_{1}+\frac{{\bf s}_1}{2}\right)...
\psi_{n'_N}\left({\bf r}_{N}+\frac{{\bf s}_N}{2}\right)
\psi^{\star}_{n''_N}\left({\bf r}_{N}-\frac{{\bf s}_N}{2}\right)...
\psi^{\star}_{n''_1}\left({\bf r}_{1}-\frac{{\bf s}_1}{2}\right)
\nonumber
\\
& = &
\sum_{n'_1}...\sum_{n'_N}\sum_{n''_1}...\sum_{n''_N}
\left\langle 0\left|\hat{c}_{n'_1}...\hat{c}_{n'_N}
\hat{c}^{\dagger}_{n''_N}...\hat{c}^{\dagger}_{n''_1}\right|0\right\rangle
e^{-\frac{1}{k_BT}(\epsilon_{n''_1}+...+\epsilon_{n''_N})}\times
\nonumber
\\
&&
\psi_{n'_1}\left({\bf r}_{1}+\frac{{\bf s}_1}{2}\right)...
\psi_{n'_N}\left({\bf r}_{N}+\frac{{\bf s}_N}{2}\right)
\psi^{\star}_{n''_N}\left({\bf r}_{N}-\frac{{\bf s}_N}{2}\right)...
\psi^{\star}_{n''_1}\left({\bf r}_{1}-\frac{{\bf s}_1}{2}\right),
\end{eqnarray}
\end{widetext}

\noindent
where $\psi_n$ indicates the $n$-th eigenstate of the confining potential. In order to better understand
how to treat the above equation let us focus on the two-particle case. Using the fermionic or the
bosonic commutation rules for the contribution containing the creation and annihilation operators,
the following identity is achieved:

\begin{equation}
\label{identity}
\Big\langle 0\Big|
\hat{c}_{n'_1}\,\hat{c}_{n'_2}\,\hat{c}^{\dagger}_{n''_2}\,\hat{c}^{\dagger}_{n''_1}
\Big| 0\Big\rangle
=
\big(\delta_{n'_1n''_1}\delta_{n'_2n''_2}\pm\delta_{n'_1n''_2}\delta_{n'_2n''_1}\big).
\end{equation}

By substituting equation (\ref{identity}) into equation (\ref{matrixelemnet}) for $N=2$, and using it
in equation (\ref{defsecond}) the following expression is obtained:

\begin{widetext}
\begin{eqnarray}
f^{(2)}_W({\bf r}_1,{\bf p}_1)
& = &
\frac{2}{h^3\mathcal Z}
\int\,d{\bf r}_2\,d{\bf p}_2\int\,d{\bf s}_1\,d{\bf s}_2
e^{-\frac{i}{\hbar}({\bf s}_1{\bf p}_1+{\bf s}_2{\bf p}_2)}
\sum_{n'_1}\sum_{n'_2}e^{-\frac{1}{k_BT}(\epsilon_{n'_1}+\epsilon_{n'_2})}
\nonumber
\\
&&
\times
\left\{
\psi_{n'_2}\left({\bf r}_{2}+\frac{{\bf s}_2}{2}\right)
\psi_{n'_1}\left({\bf r}_{1}+\frac{{\bf s}_1}{2}\right)
\psi^{\star}_{n'_1}\left({\bf r}_{1}-\frac{{\bf s}_1}{2}\right)
\psi^{\star}_{n'_2}\left({\bf r}_{2}-\frac{{\bf s}_2}{2}\right)
\right.
\nonumber
\\
&&
\pm
\left.
\psi_{n'_1}\left({\bf r}_{2}+\frac{{\bf s}_2}{2}\right)
\psi_{n'_2}\left({\bf r}_{1}+\frac{{\bf s}_1}{2}\right)
\psi^{\star}_{n'_1}\left({\bf r}_{1}-\frac{{\bf s}_1}{2}\right)
\psi^{\star}_{n'_2}\left({\bf r}_{2}-\frac{{\bf s}_2}{2}\right)
\right\},
\end{eqnarray}

\noindent
then, performing the integrals, we finally get:

\begin{eqnarray}
f^{(2)}_W({\bf r}_1,{\bf p}_1)
& = &
\frac{2}{\mathcal Z}
\sum_{n'_1}\sum_{n'_2}
e^{-\frac{1}{k_BT}(\epsilon_{n'_1}+\epsilon_{n'_2})}
f_{W_{n'_1}}({\bf r}_1,{\bf p}_1)
\nonumber
\\
&&
\pm
\frac{2}{\mathcal Z}
\sum_{n'_1}\sum_{n'_2}
e^{-\frac{1}{k_BT}(\epsilon_{n'_1}+\epsilon_{n'_2})}
\delta_{n'_2n'_1}
\int\,d{\bf s}_1
e^{-\frac{i}{\hbar}{\bf s}_1{\bf p}_1}
\psi_{n'_2}\left({\bf r}_{1}+\frac{{\bf s}_1}{2}\right)
\psi^{\star}_{n'_1}\left({\bf r}_{1}-\frac{{\bf s}_1}{2}\right)
\nonumber
\\
& = &
\frac{2}{\mathcal Z}
\sum_{n'_1}
\left\{\sum_{n'_2}
e^{-\frac{1}{k_BT}(\epsilon_{n'_1}+\epsilon_{n'_2})}
\pm
e^{-\frac{1}{k_BT}2\epsilon_{n'_1}}
\right\}
f_{W_{n'_1}}({\bf r}_1,{\bf p}_1),
\end{eqnarray}
\end{widetext}

\noindent
where $f_{W_{n_1}}({\bf r}_1,{\bf p}_1)$ indicates the WF of the $n^{th}_1$ eigenstate of the
confined potential.

This expression can be written in a more compact form by identifying the bosonic and the fermionic
cases:

\begin{equation}
f^{(2)}_{W_{bosons}}({\bf r}_1,{\bf p}_1)=
\frac{2}{\mathcal{Z}}
\sum_{n_1}\sum_{n_2}
e^{-\frac{\epsilon_{n_1}+\epsilon_{n_2}}{k_BT}}
f_{W_{n_1}}({\bf r}_1,{\bf p}_1),
\end{equation}

\begin{equation}
f^{(2)}_{W_{fermions}}({\bf r}_1,{\bf p}_1)=
\frac{2}{\mathcal{Z}}
\sum_{n_1}\sum_{n_2\neq n_1}
e^{-\frac{\epsilon_{n_1}+\epsilon_{n_2}}{k_BT}}
f_{W_{n_1}}({\bf r}_1,{\bf p}_1).
\end{equation}

The generalization to the $N$-particle system is straightforward and gives:

\begin{eqnarray}
f^{(N)}_{W_{bosons}}({\bf r}_1,{\bf p}_1)
& = &
\frac{N}{\mathcal{Z}}
\sum_{n_1,n_2,...,n_N}
e^{-\frac{\epsilon_{n_1}+\epsilon_{n_2}+...+\epsilon_{n_N}}{k_BT}}
\nonumber
\\
&&
\times
f_{W_{n_1}}({\bf r}_1,{\bf p}_1),
\end{eqnarray}

\begin{eqnarray}
\label{equfin}
f^{(N)}_{W_{fermions}}({\bf r}_1,{\bf p}_1)
& = &
\frac{N}{\mathcal{Z}}
\sum_{n_1\neq n_2...\neq n_N}
e^{-\frac{\epsilon_{n_1}+\epsilon_{n_2}+...+\epsilon_{n_N}}{k_BT}}
\nonumber
\\
&&
\times
f_{W_{n_1}}({\bf r}_1,{\bf p}_1).
\nonumber
\\
\end{eqnarray}

Let us note that in the above expression for the fermionic case the following sum appears:

\begin{equation}
\frac{1}{\mathcal{Z}}
\sum_{n_2\neq n_1}
...
\sum_{n_N\neq n_{N-1}\neq...\neq n_1}
e^{-\frac{\epsilon_{n_1}+\epsilon_{n_2}+...+\epsilon_{n_N}}{k_BT}}.
\end{equation}

\noindent
In the limit of large $N$ and of an infinite number of allowed states with continuous energies
the above term gives the Fermi function evaluated at an energy value $\epsilon_{n_1}$ \cite{Reif}.

\subsection{Infinite Square Potential Well}

\begin{figure}
\vspace{1cm}
\centering
\includegraphics[width=7.5cm,angle=0]{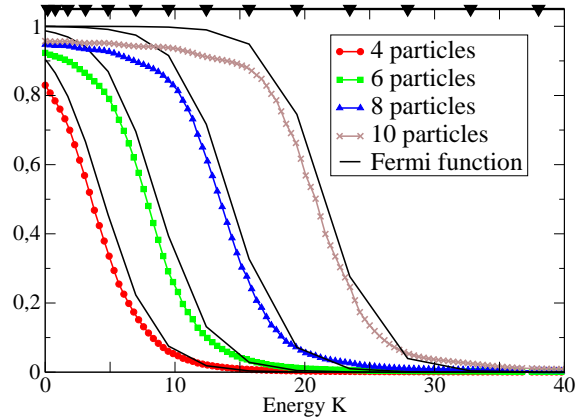}
\caption{\label{BUCAf} Average values of the points corresponding to the same energy inteval of the
single-particle reduced WF of a system of $N$ particles at thermal equilibrium at a temperature of
$2$ K in an 1D infinite square well potential. Since the energy depends only upon the momenta of the
particles, the above average corresponds to the integral over the position variable ($x$). The width
of the well has been fixed to 150 nm. In the upper part of the figure the black triangles indicate
the energies corresponding to the eigenstates of the well.}
\end{figure}

An inifite square well potential in one dimension has been investigated at a temperature of $T=2$ K.
The single-particle WF has been evaluated for $N=4,6,8$ and $10$ by means of equation (\ref{equfin}).
Then the average values of the points of the WF corresponding to energy interval $\epsilon$,
$\epsilon+\delta\epsilon$ have been plotted as a function of $\epsilon$, in Fig. \ref{BUCAf}.
Since the energy depends only upon the Wigner momentum of the particle ($p^2/2m$), the above average
corresponds to the integral of the single-particle reduced WF with respect to the position variable
($x$). In our simulations the width of the well has been kept contant to a value of 150 nm.

A comparison between our curves and the Fermi functions is obtained by evaluating the chemical
potentials $\mu$ for $N=4,6,8$ and $10$ from a numerical solution of the equation:

\begin{equation}
\label{chempot}
N=\sum_{n=1}^{\infty}\frac{1}{e^{\frac{\epsilon_n-\mu}{k_BT}}+1}.
\end{equation}

\noindent
The curves in Fig. \ref{BUCAf} show a very good agrement between the Fermi function and the average
of the WF for any number of particles.

\begin{figure}
\vspace{1cm}
\centering
\includegraphics[width=7.5cm,angle=0]{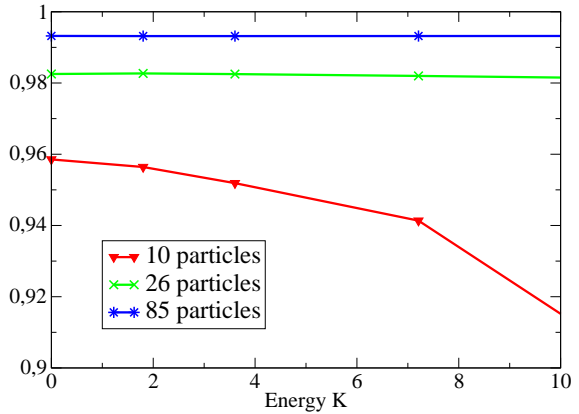}
\caption{\label{ditail} Average values of the points corresponding to the same energy of the
single-particle reduced WF of a system of $N$ particles at thermal equilibrium at a temperature of
$2$ K in a 1D infinite square well potential. The width of the well has been fixed to 150 nm and the
number of particles increased from 10 to 85.}
\end{figure}

It is worth noting that, as expected, the agreement between the averages of WFs and the Fermi
distributions is higher as the number of particles increases. However in Fig. \ref{BUCAf} even in
the case of $10$ particles the value of the WF's average corresponding to the point with $\epsilon=0$
do not reach the maximal value of $1$. For this reason we have plotted in Fig. \ref{ditail} the
system in more ditail in the energy range from $0$ to $10$ K for a higher number of particles.
As before we simulate a well $150$ nm wide with an electron gas at a temperature of $2$ K, in this
case, however, the number of simulated particles is increased to $85$. Fig. \ref{ditail} shows that
when the number of particles increases the value corresponding to $\epsilon=0$ approaches $1$ .

\subsection{Harmonic Potential}

\begin{figure}
\vspace{1cm}
\centering
\includegraphics[width=7.5cm]{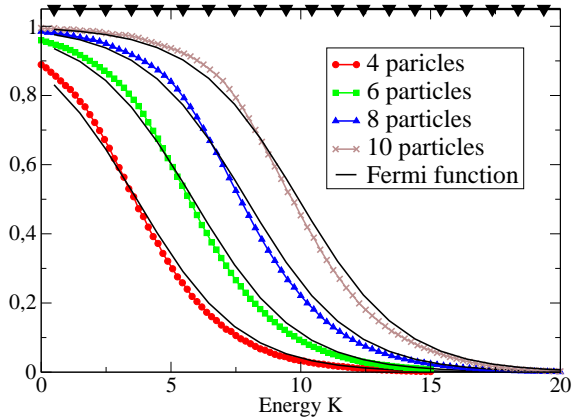}
\caption{\label{PAf} Average values of the points corresponding to the same energy of the reduced
single-particle WF in a harmonic potential with spring constant $k=1.54\times10^8$ [$kg/sec^2$].
The system in the case of $N=4,6,8,10$ particles is studied at thermal equilibrium at a temperature
of $2$ K. The curves clearly tend to the Fermi-Dirac distribution. In the upper part of the figure
the black triangles indicate the energies corresponding to the eigenstates of the harmonic potential.}
\end{figure}

As a second example, we have studied a one-dimensional harmonic potential. Equation (\ref{equfin})
has been evaluated for different numbers of particles ($4,6,8$ and $10$) at a temperture of $T=2$ K.
The average over the points of the WF belonging to the same energy interval ($p^2/2m+\frac{1}{2}kx^2$,
where $m$ is the mass of any particle in the system) are plotted in Fig \ref{PAf}. The Fermi function
with the chemical potential given by equation (\ref{chempot}) is clearly approached by the
corresponding average of the WF.

\subsection{Effect of energy separation between levels}

\begin{figure}
\vspace{1cm}
\centering
\includegraphics[width=7.5cm]{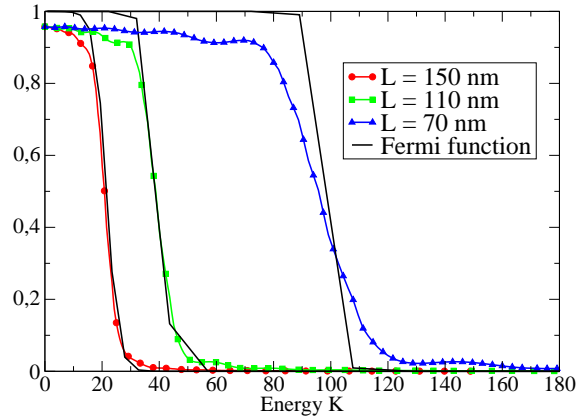}
\caption{\label{Lmodulation} Average values of the points corresponding to the same energy of the
reduced single-particle WF in a 1D infinite square potential well. In a system with 10 particles at
a temperature of $2$ K, the width of the well has been reduced from $150$ nm to $70$ nm, that corresponds
to an energy gap decrease, from the ninth to the tenth energy level, from $3.8$ to $17.1$ K.
When the energy spacing between the levels is bigger than the thermal energy, the particles distribution
deviates form a Fermi function.}
\end{figure}

\begin{figure}
\vspace{1cm}
\centering
\includegraphics[width=7.5cm]{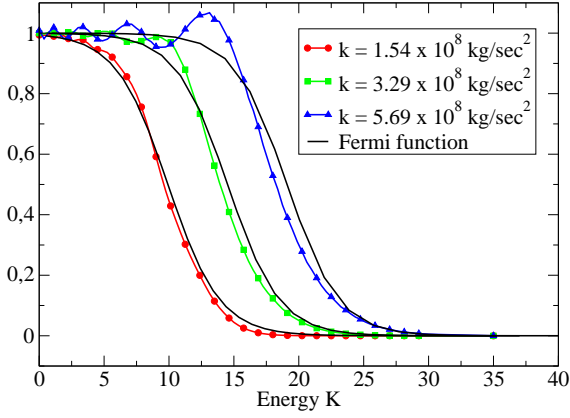}
\caption{\label{BCmodulation} Average values of the points corresponding to the same energy interval
of the reduced single-particle WF in a harmonic potential. In the case of 10 particles the bound
constant has been varied from $k=1.54\times10^8$ [$kg/sec^2$] to $k=5.69\times10^8$ [$kg/sec^2$]
corresponding to an increase of the spacing between the energy levels from 1 to 2 K. When the gap
between the energy levels increases and becomes bigger than $2$ K, the gas temperature, the particles
distribution shows an oscillating behaviuor superimposed to the Fermi-like shape.}
\end{figure}

It is possible to study how the particle distributions change when the dimension of the well or
the strength of the harmonic potential are varied in the case of both the infinite square well
potential and the harmonic potential. When the width of the well decreases or the strength of the
harmonic potential increases, the spacing between the allowed energy levels increases and an
oscillating behaviour shows up in the curves (see Fig.s~\ref{Lmodulation},~\ref{BCmodulation}). Our
calculations have been performed in the case of a system with $10$ particles at a temperature of
$2$ K. When the width of the well is decreased from $150$ nm to $70$ nm, the energy gap between
the ninth and the tenth energy level increases from $3.8$ to $17.1$ K. In the case of the harmonic
potential the strength of the force constant is varied from $1.54\times10^8$ to $5.69\times10^8$
kg/sec$^2$ leading to an increase of the distance between the energy levels from $1$ to $2$ K. The
simulations show that such oscillations get more and more evident as the ratio between
the spacing of the energy levels and $k_BT$ becomes greater. Under these conditions the picture of
a continuous spectrum of energies breaks down.

\section{Transport Equation}

The dynamical equation for the single-particle WF is derived by differentiating the definition of
the WF itself:

\begin{equation}
\frac{\partial}{\partial t}f_W({\bf r},{\bf p},t)
\hspace{-0.05cm}
=
\hspace{-0.15cm}
\int d{\bf s}e^{-\frac{i}{\hbar}{\bf sp}}\frac{\partial}{\partial t}
\left[\psi\left({\bf r}+\frac{{\bf s}}{2},t\right)
\psi^{\star}\left({\bf r}-\frac{{\bf s}}{2},t\right)\right].
\end{equation}

By means of the  Schr\"odinger equation it is possible to evaluate the time derivative of the product
of the two wave functions and to obtain the dynamical equation for the WF \cite{Jacoboni1}:

\begin{eqnarray}
\frac{\partial}{\partial t}f_W({\bf r},{\bf p},t)
& = &
-\frac{{\bf p}}{m}\nabla f_W({\bf r},{\bf p},t)
\nonumber
\\
&&
+\frac{1}{h^3}\int\,d{\bf p}'{\mathcal V}_W({\bf r},{\bf p}-{\bf p}')f_W({\bf r},{\bf p}',t),
\nonumber
\\
\end{eqnarray}

\noindent
where ${\mathcal V}_W$ is the interaction kernel for an external potential $V({\bf r})$. Note that
the interaction term, given by:

\begin{equation}
{\mathcal V}_W({\bf r},{\bf p})
\hspace{-0.1cm}
=
\hspace{-0.1cm}
\frac{1}{i\hbar}\int
d{\bf s}\,e^{-\frac{i}{\hbar}{\bf ps}}
\left[V
\hspace{-0.1cm}
\left({\bf r}+\frac{{\bf s}}{2}\right)
\hspace{-0.1cm}
-
\hspace{-0.1cm}
V
\hspace{-0.1cm}
\left({\bf r}-\frac{{\bf s}}{2}\right)\right],
\end{equation}

\noindent
depends on the values of $V$ at points different from ${\bf r}$. However, while the non-locality of
${\mathcal V}_W$ extends to infinity, its effect on the electron dynamics has to be considered only
up to regions where the electron correlation is different from zero.

\subsection{Electron-Electron Scattering}

Let us study the transport equation for electron-electron scattering. In the case where no phonons
nor external forces are present, the potential $V({\bf r}_1,{\bf r}_2...{\bf r}_N)$ is given by the
Coulomb interaction, the transport equation reads:

\begin{widetext}
\begin{eqnarray}
\label{eescattering}
\frac{\partial}{\partial t}f_W({\bf r}_1,{\bf p}_1,...,{\bf r}_N,{\bf p}_N,t)
& = &
-\sum_l\frac{{\bf p}_l}{m}\nabla_{{\bf r}_l}
f_W({\bf r}_1,{\bf p}_1,...,{\bf r}_l,{\bf p}_l,...,{\bf r}_N,{\bf p}_N,t)
\nonumber\\
&&
+\frac{1}{\hbar^3}\sum_i\sum_j\int\,d{\bf p}'_id{\bf p}'_j\delta(\Delta {\bf p}_i+\Delta {\bf p}_j)
V_W(|{\bf r}_i-{\bf r}_j|,\Delta {\bf p}_i-\Delta {\bf p}_j)
\nonumber\\
&&
\times
f_W({\bf r}_1,{\bf p}_1,...,{\bf r}_i,{\bf p}'_i,...,{\bf r}_j,{\bf p}'_j,...{\bf r}_N,{\bf p}_N,t),
\nonumber
\\
\end{eqnarray}
\end{widetext}

\noindent
where $V_W$ is the potential kernel of the Wigner equation and $\Delta {\bf p}={\bf p}-{\bf p}^{\prime}$.
As done before, in order to get a better understanding of the above equation, the kernel that describes
the electron-electron interaction is studied for $N=2$:

\begin{eqnarray}
V_W({\bf r}_1,{\bf r}_2,{\bf p}_1,{\bf p}_2)
& = &
\frac{1}{i\hbar}\int\,d{\bf s}_1\,d{\bf s}_2
e^{-\frac{i}{\hbar}({\bf p}_1{\bf s}_1+{\bf p}_2{\bf s}_2)}
\nonumber
\\
&&
\hspace{-3cm}
\times
\Big[V\Big({\bf r}_1+\frac{{\bf s}_1}{2},{\bf r}_2+\frac{{\bf s}_2}{2}\Big)
-
V\Big({\bf r}_1-\frac{{\bf s}_1}{2},{\bf r}_2-\frac{{\bf s}_2}{2}\Big)\Big].
\nonumber
\\
\end{eqnarray}

\noindent
Since the Coulomb interaction depends only upon the distance between the two particles it is usefull
to re-write the above equation using the new variables ${\bf x}={\bf r}_1-{\bf r}_2$,
${\bf s}={\bf s}_1-{\bf s}_2$, and
${\bf s}'=({\bf p}_1{\bf s}_2+{\bf p}_2{\bf s}_1)/({\bf p}_1+{\bf p}_2)$:

\begin{widetext}
\begin{eqnarray}
V_W({\bf r}_1,{\bf r}_2,{\bf p}_1,{\bf p}_2)
& = &
\frac{1}{i\hbar}\int\,d{\bf s}\,d{\bf s}'
e^{-\frac{i}{\hbar}({\bf p}_1+{\bf p}_2){\bf s}'}
e^{-\frac{i}{\hbar}({\bf p}_1-{\bf p}_2){\bf s}}
\Big[V\Big({\bf x}+\frac{{\bf s}}{2}\Big)-V\Big({\bf x}-\frac{{\bf s}}{2}\Big)\Big]
\nonumber
\\
& = &
\frac{\hbar^3}{i\hbar}\delta({\bf p}_1+{\bf p}_2)\int\,d{\bf s}
e^{-\frac{i}{\hbar}({\bf p}_1-{\bf p}_2){\bf s}}
\Big[V\Big({\bf x}+\frac{{\bf s}}{2}\Big)-V\Big({\bf x}-\frac{{\bf s}}{2}\Big)\Big]
\nonumber
\\
& = &
\hbar^3\delta({\bf p}_1+{\bf p}_2)V_W({\bf r}_1-{\bf r}_2,{\bf p}_1-{\bf p}_2).
\end{eqnarray}
\end{widetext}

Thus the factor $\delta(\Delta {\bf p}_i+\Delta {\bf p}_j)$ appearing in equation (\ref{eescattering})
represents the constrain for the total momentum consevation while the difference
$(\Delta {\bf p}_i-\Delta {\bf p}_j)$ indicates that the interaction depends only upon the momentum
transfer between particle $i$ and $j$.

When the single-particle reduced WF in the case of $N$ particles is evaluated, equation
(\ref{eescattering}) reads:

\begin{eqnarray}
\frac{\partial}{\partial t}f^{(N)}_W({\bf r},{\bf p},t)
& = &
-\frac{{\bf p}}{m}\nabla_{\bf r}f^{(N)}_W({\bf r},{\bf p},t)
\nonumber\\
&&
\hspace{-0.35cm}
+\frac{1}{\hbar^3}\int\,d\varrho\,d{\bf p}_{\varrho}\int\,d{\bf p}'V_W(|{\bf r}-\varrho|,2\Delta {\bf p})
\nonumber\\
&&
\hspace{-0.35cm}
\times
f^{(N)}_W({\bf r},{\bf p}',\varrho,{\bf p}_{\varrho},t),
\end{eqnarray}

\noindent
where ${\bf r}$ and ${\bf p}$ are the position and the Wigner momentum of the considered particle
and, $\varrho$ and ${\bf p}_{\varrho}$ indicate the position coordinates of one of the remaining $N-1$
particles. It should be noticed that all the particles are interacting with each other, but, due to their
indistinguishability, all the contributions are identical and sum up to balance the factorials appearing
in equation (\ref{reduced}). The above expression shows that the transport equation for the reduced
single-particle WF depends on the reduced two-particle WF. When the transport equation for the reduced
two-particle WF is evaluated, the electron-electron interaction term depends upon the three-particle
reduced WF and so on for the transport equation for the other reduced WFs. It is the Wigner picture of
the BBGKY hierarchy.

When the WF is written in terms of antisimmetric single-particle wave functions, as we have seen in
equation (\ref{overlapping}), two types of contributions can be identified. It is then possible to
study how the transport equation reads when only the contributions due to non overlapping wave functions
are considered:

\begin{widetext}
\begin{eqnarray}
\frac{\partial}{\partial t}f^{(N)}_W({\bf r},{\bf p},t)
& = &
\frac{\partial}{\partial t}\sum_if^{(N)}_{W_i}({\bf r},{\bf p},t)
\nonumber\\
& = &
-\frac{{\bf p}}{m}\nabla_{\bf r}f^{(N)}_W({\bf r},{\bf p},t)
+\frac{1}{\hbar^3}\sum_i\int\,d{\bf p}'\,d{\bf \varrho}\,
V_W(|{\bf r}-{\bf \varrho}|,2\Delta{\bf p})\left[\sum_{j\neq i}|\psi_j({\bf \varrho})|^2\right]
\nonumber\\
&&
\times
f^{(N)}_{W_i}({\bf r},{\bf p}',t).
\nonumber\\
\end{eqnarray}
\end{widetext}

\noindent
Besides a Liouvillian contribution, an interaction term appears where each one-particle contribution
interacts with all the others as in the Hartree approximation. In the case of overlapping wave functions
also the other kind of contributions (as studied in equation (\ref{overlapping})) must be considered,
and the exchange term is restored.

\subsection{Electron-Phonon Scattering for the 2-Electrons WF}

The e-ph interaction term for the dynamical equation for two electrons is made out of eight
terms as follows:

\begin{widetext}
\begin{eqnarray}
\label{2e-ph}
\frac{\partial}{\partial t}f_W\Bigg|_{ep}
\hspace{-0.15cm}
& = &
\hspace{-0.15cm}
\sum_{{\bf q}'}F({\bf q}')
\nonumber
\\
&&
\hspace{-0.20cm}
\times
\hspace{-0.05cm}
\left\{e^{i({\bf q}'{\bf r}_1-\omega_{q'}(t-t_0))}\sqrt{n_{q'}+1}
\right.
f_W\left(
{\bf r}_1,{\bf p}_1-\frac{\hbar{\bf q}'}{2},{\bf r}_2,{\bf p}_2,\{...,n_{q'}+1,...\},\{n'_{q}\},t
\right)
\nonumber
\\
&&
\hspace{-0.20cm}
-e^{-i({\bf q}'{\bf r}_1-\omega_{q'}(t-t_0))}\sqrt{n_{q'}}
\,f_W\left(
{\bf r}_1,{\bf p}_1+\frac{\hbar{\bf q}'}{2},{\bf r}_2,{\bf p}_2,\{...,n_{q'}-1,...\},\{n'_{q}\},t
\right)
\nonumber
\\
&&
\hspace{-0.20cm}
+e^{-i({\bf q}'{\bf r}_1-\omega_{q'}(t-t_0))}\sqrt{n'_{q'}+1}
\,f_W\left(
{\bf r}_1,{\bf p}_1-\frac{\hbar{\bf q}'}{2},{\bf r}_2,{\bf p}_2,\{n_q\},\{...,n'_{q'}+1,...\},t
\right)
\nonumber
\\
&&
\hspace{-0.20cm}
-\left.
e^{i({\bf q}'{\bf r}_1-\omega_{q'}(t-t_0))}\sqrt{n'_{q'}}
\,f_W\left(
{\bf r}_1,{\bf p}_1+\frac{\hbar{\bf q}'}{2},{\bf r}_2,{\bf p}_2,\{n_q\},\{...,n'_{q'}-1,...\},t
\right)
+o.p.\right\},
\nonumber
\\
\end{eqnarray}
\end{widetext}

\noindent
where ${\bf r}_1,{\bf p}_1,{\bf r}_2,{\bf p}_2$ are the Wigner phase space coordinates of the two
particles. In the above equation $o.p.$ stands for $other$ $particle$ and indicates the four terms
where ${\bf r}_2$ replaces ${\bf r}_1$ in the exponential factors and ${\bf p}_2$ undergoes a
variation of $\hbar{\bf q}'/2$ while ${\bf p}_1$ remains unchanged.

The eight terms appearing on the $r.h.s.$ of the above equation have simple physical interpretations:
the e-ph interaction occurs as emission or absorpion of a quantum of any mode ${\bf q}$ and this
may appear in the state on the left or on the right of the bilinear expression that defines the WF.
Each elementary interaction or $vertex$ changes only one of the two sets of variables of the WF; more
precisely, one of the occupation numbers $n_q$ is changed by unity and one of the electron momenta
is changed by half of the phonon momentum.

In analogy with the Chambers formulation \cite{chambers} of the classical kinetic equation it is possible
to introduce new variables $({\bf r}_i^*,{\bf p}_i^*,t^*)$ that allow us to obtain an integral form
of the dynamical equation for the WF. This integral equation is in a closed form and can be solved by
iteratively substituting it into itself, leading to what is known as its Neumann expansion.

Equation (\ref{2e-ph}) gives $8$ terms for the contribution of the first order of the Neumann
expansion, $64$ terms for the contribution of the second order and so on for the higher order
terms. In order to obtain meaningful physical quantities, however, the trace over the phonon modes
must be performed, leading to a vanishing contribution for each term corresponding to an odd order
in the Neumann expansion \cite{Pascoli}. Only terms with an even number of verteces give dyagonal
(in the phonon modes) contributions different from zero.

As stated before, the second order in the Neumann expansion gives $64$ terms. Among these, $32$ yield
contributions dyagonal in the phonon modes and survive to the trace operation, $16$ terms refer to one
particle and $16$ to the other. For each particle $8$ terms are the complex conjugate of the other $8$
and can be summed together leading to $8$ contributions for each particle. Among these it is possible to
recognize four standard interactions undergone by each particle: real emission; real absorption; virtual
emission and virtual absorption.

The main difference to the single particle case lies in the eight (four for each particle)  remaining
terms. In the two-particle case the phonon occupation number can be changed, in the first or second
set of values in the arguments of the WF, not just by one electron loosing (gaining) a Wigner momentum
equal to half the phonon momentum in each of the two verteces \cite{Pascoli} but also by the action
of two electrons. One electron looses or gains half phonon momentum in the first vertex and another
electron looses or gains half phonon momentum in the second vertex. Since we are dealing with identical
particles we don't know which electron interact with the phonon bath in the first or in the second
vertex. These four terms are real or virtual emissions or absorptions where the interaction with the
phonon bath is shared between the two electrons.

It should be recalled that the ${\bf p}$ variable of the WF is obtained as a linear combination of two
electron momenta. A specific value $\tilde{{\bf p}}$ is obtained as
$\tilde{{\bf p}}=\hbar\,({\bf k}_1+{\bf k}_2)/2$ where ${\bf k}_1$ and ${\bf k}_2$ range from $-\infty$
to $+\infty$. For this reason the Wigner momentum undergoes a change corresponding to half of the phonon
momentum at each interaction vertex.

\begin{figure}  
\includegraphics[width=7.0cm]{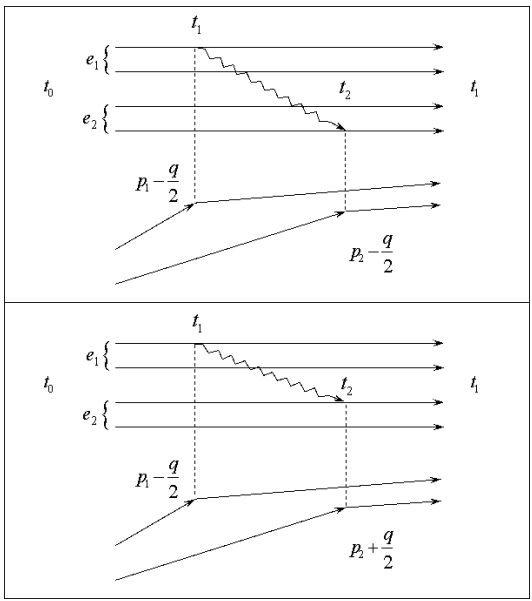}
\caption{\label{eephpath}
Real emission of a phonon mode in mutual partecipation by two electrons.
One electron $e_1$ changes its Wigner momentum by half of the
phonon momentum at time $t=t_1$ and the other electron $e_2$ does the
same at a later time $t=t_2$ (upper bow). Virtual emission where one electron
lose a Wigner momentum equal to half the momentum of the phonon while the other
electron gains the same amount (lower bow).}
\end{figure}

In Fig. \ref{eephpath}, by means of the Keldysh-diagram formalism, two of the co-participated
graphs are shown: a real phonon emission due to electron-electron cooperation and a virtual emission
respectively. In each case both the Keldysh diagram representing the transition and the corresponding
Wigner path \cite{Pascoli} undergone by the two electrons are shown. Since the Keldysh diagrams are in
the density matrix (DM) representation, four time-lines appear, two for each electron. The first and
the third lines correspond to the first wave function of the DM while the other two correspond to the
second wave function. In the upper box at time $t=t_1$ one electron in the first wave function of
the DM emits a phonon and at time $t=t_2$ another electron in the second wave function emits the same
phonon.

In the lower box of Fig. \ref{eephpath} at time $t=t_2$ an electron absorbs the phonon emitted at
$t=t_1$, corresponding to a virtual emission.

\section{Conclusion}

We have developed a model based on the WF formalism that allows to introduce the
symmetry  effect in a system where electrons interact with each other and with
the phonon bath. We have shown how this formalism can be usefull by applying it
to different situations: the study of the electron-electron scattering, the study of
the thermal distribution of N particles in confining potentials, and the study of
the two-electron dynamics in the presence of electron-phonon scattering. We have also
shown that with the WF it is possible to reproduce a Fermi like distribution defining
the energy in the Wigner phase-space. 

\begin{acknowledgments}
This work has been partially supported by the U.S.Office of Naval Research (contract No.
N00014-03-1-0289).
\end{acknowledgments}

\bibliography{2particles}

\begin{thebibliography}{1}
\bibitem{Wigner} E. Wigner, Phys. Rev., 40, 749-759, 1932
\bibitem{Frensley} W. Frensley, Rev Mod. Phys., 62, 745-791, 1990.
\bibitem{Nedjalkov} M. Nedjalkov et al., Phys. Rev. B, 70, 115319-115335, 2004.
\bibitem{Jacoboni1} C. Jacoboni et al., Rep. Prog. Phys., 67, 1033-1071, 2004.
\bibitem{Carruters} P. Carruters et al., Rew. Mod. Phys., 55, 245-285, 1983.
\bibitem{Rossi} F. Rossi et al., Rev. Mod. Phys., 74, 895-950, 2002.
\bibitem{Keldysh} L.V. Keldysh, Zh. Eksp Teor. Fiz 47, 1515-1527, 1964, Sov. Phys. JEPT 20, 1018, 1965.
\bibitem{Pascoli} M. Pascoli et al., Phys. Rev. B., 58, 3503-3506, 1998.
\bibitem{Kirkwood} J.G. Kirkwood, Phys. Rev., 44, 31-37, 1933
\bibitem{Harper} C. Harper, Am. J. Phys., 42, 396-399, 1974
\bibitem{Uhlenbeck} G.E. Uhlenbeck et al., Phys. Rev., 41, 79-89, 1932
\bibitem{Hillery} M. Hillery et al., Phys. Rep., 106, 121-167, 1984.
\bibitem{Imre} K. Imre et al., J. Math. Phys., 8, 1097-1108, 1967.
\bibitem{iwce10} E. Cancellieri et al., J. Comp. Electron., 3, 411-415, 2004.
\bibitem{Reif} F. Reif, McGraw-Hill International Book Company, ``Statistical and Thermal Physics'', 1965.
\bibitem{chambers} R.G. Chambers, Proc. Phys. Soc. A, 65,458,1952.
\end{thebibliography}

\end{document}